\documentclass[letterpaper]{article} % DO NOT CHANGE THIS
\usepackage{aaai25}  % DO NOT CHANGE THIS
\usepackage{times}  % DO NOT CHANGE THIS
\usepackage{helvet}  % DO NOT CHANGE THIS
\usepackage{courier}  % DO NOT CHANGE THIS
\usepackage[hyphens]{url}  % DO NOT CHANGE THIS
\usepackage{graphicx} % DO NOT CHANGE THIS
\def\UrlFont{\rm}  % DO NOT CHANGE THIS
\usepackage{natbib}  % DO NOT CHANGE THIS AND DO NOT ADD ANY OPTIONS TO IT
\usepackage{caption} % DO NOT CHANGE THIS AND DO NOT ADD ANY OPTIONS TO IT
\usepackage{algorithm}
\usepackage{algorithmic}
\usepackage{multirow}
\usepackage{dsfont}
\usepackage{tabularray}
\usepackage{amsmath}

\usepackage[frozencache,cachedir=.]{minted}

\usepackage{fancyhdr} 
\usepackage{pifont}
\newcommand{\cmark}{\text{\ding{51}}}
\newcommand{\xmark}{\text{\ding{55}}}

\usepackage{newfloat}
\usepackage{listings}
\DeclareCaptionStyle{ruled}{labelfont=normalfont,labelsep=colon,strut=off} % DO NOT CHANGE THIS
\floatstyle{ruled}
\newfloat{listing}{tb}{lst}{}
\floatname{listing}{Listing}
\title{ScriptSmith: A Unified LLM Framework for Enhancing IT Operations via Automated Bash Script Generation, Assessment, and Refinement}

\author {
    Oishik Chatterjee\textsuperscript{\rm 1},
    Pooja Aggarwal\textsuperscript{\rm 1},
    Suranjana Samanta\textsuperscript{\rm 1},
    Ting Dai\textsuperscript{\rm 1},
    Prateeti Mohapatra\textsuperscript{\rm 1},
    Debanjana Kar\textsuperscript{\rm 1},
    Ruchi Mahindru\textsuperscript{\rm 1},
    Steve Barbieri\textsuperscript{\rm 2},
    Eugen Postea\textsuperscript{\rm 2},
    Brad Blancett\textsuperscript{\rm 2},
    Arthur De Magalhaes\textsuperscript{\rm 2}
}
\affiliations {
    \textsuperscript{\rm 1}IBM Research\\
    \textsuperscript{\rm 2}IBM Software\\
    \{oishik.chatterjee, ting.dai, debanjana.kar1\}@ibm.com, \{aggarwal.pooja, pramoh01, suransam\}@in.ibm.com,\\ \{blancett, rmahindr,barbier\}@us.ibm.com, \{epostea, arthurdm\}@ca.ibm.com
}

\begin{document}

\maketitle

\begin{abstract}
In the rapidly evolving landscape of site reliability engineering (SRE), the demand for efficient and effective solutions to manage and resolve issues in site and cloud applications is paramount. This paper presents an innovative approach to action automation using large language models (LLMs) for script generation, assessment, and refinement. By leveraging the capabilities of LLMs, we aim to significantly reduce the human effort involved in writing and debugging scripts, thereby enhancing the productivity of SRE teams. Our experiments focus on Bash scripts, a commonly used tool in SRE, and involve the CodeSift~\cite{codesift2024} dataset of $100$ tasks and the InterCode~\cite{intercode2023} dataset of $153$ tasks. The results show that LLMs can automatically assess and refine scripts efficiently,  reducing the need for script validation in an execution environment. Results demonstrate that the framework shows an overall improvement of $7-10\%$ in script generation.
\end{abstract}

\section{Introduction}

Modern IT's growing complexity in multi-cloud environments creates challenges for SREs, as they strive to ensure systems operate efficiently. Organizations face the challenge of managing a growing number of incidents and outages across a diverse range of technologies and complex environments. Automation is essential to improve IT operations efficiency and reduce incident resolution time. 
A typical Incident Remediation pipeline (Figure ~\ref{fig:arch}) consists of (1) Root cause diagnosis which creates an incident report with probable root cause, (2) Action Recommendation that provides actionable recommendations, and (3) Action Automation where action recommendation outputs are transformed into scripts that can be executed to resolve the incidents. %An example of an incident flow is shown in Figure ~\ref{fig:Intro_Example}. %For instance, a SRE has been notified of a sudden increase in the number of erroneous calls. An event/incident is created and the SRE uses his prior and experential knowledge to solve the incident.

%With recent advances in Large Language Models (LLMs) helping with a significant leap in code generation tasks, utilizing models that translate natural language actionable recommendations into code or scripts can significantly reduce the manual effort involved in writing and debugging, leading to increased productivity for engineers. 
From our experience, we have seen that domain-specific scripting languages like Bash and PowerShell are commonly used in IT operations (ITOPs) for action tasks. Recent advances in Large Language Models (LLMs) have made it easier to turn natural language recommendations into script. This reduces the manual work of writing and debugging, boosting productivity for SREs.

Existing work on code benchmarks, generation, and assessment~\cite{nl2bash2018,chen2021evaluatinglargelanguagemodels,mbpp2021,intercode2023} focuses on runtime testing by evaluating code against predefined input-output specifications. These benchmarks typically assume a pre-configured environment and measure how well the generated code performs specific functions under these conditions.
For system-related scripts, two major challenges arise. First, an execution environment for testing the scripts may not always be available. Second, the values for parameters in the generated script may vary due to dependencies on the environmental context.
For example, if a task is to identify available system memory, the value is dynamic and changes over time. This variability complicates the verification of script correctness in traditional execution environments. To address this challenge, we design a framework for automatic bash script generation, assessment, and refinement that does not depend on the execution environment.

Our contributions can be summarized as follows:
\begin{itemize}
    \item \textbf{\textit{Execution Free Framework:}} We propose \textbf{\textit{ScriptSmith}}, a novel reference and execution-free automated bash script generation and refinement framework.
    \item \textit{\textbf{Uncover Gaps in LLMs and Identify Opportunities:}} We conduct extensive experiments with various LLMs and prompting techniques, revealing gaps and opportunities for improvement in this field.
    \item \textbf{\textit{Demonstrate Framework Efficacy with Human Assessment:}} We evaluate our framework through a user study with domain experts to ensure its readiness for deployment with confidence.
\end{itemize}

\begin{figure*}[htb]
\centering
\includegraphics[width=0.85\linewidth]{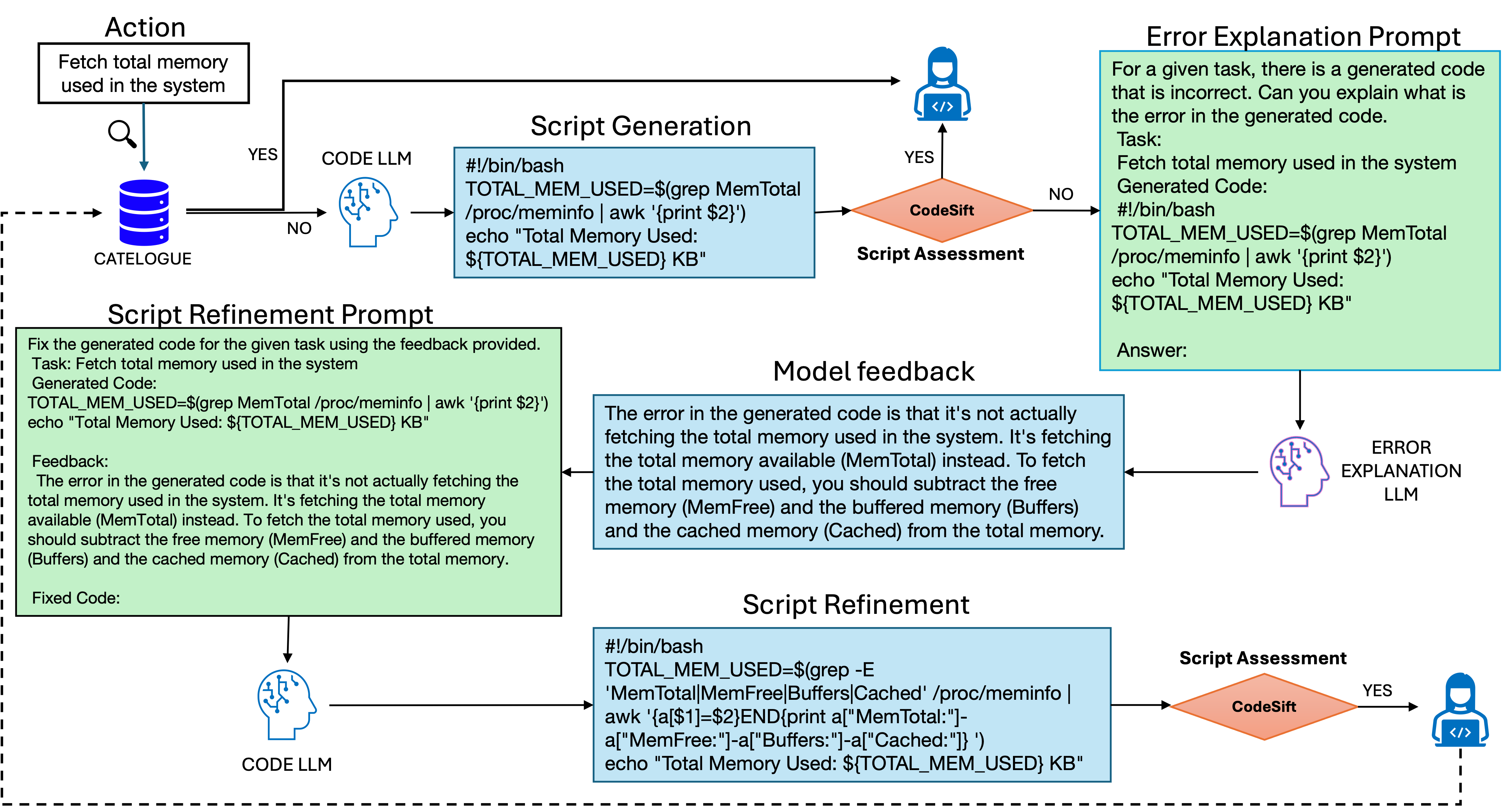}
\caption{Framework of our proposed method of ScriptSmith for generating bash scripts for incident remediation using LLMs.}
\label{fig:framework}
\end{figure*}

\iffalse
 \begin{figure}[!htp]
      \centering
      \includegraphics[width=1.0\linewidth]{CameraReady/LaTeX/figs/Intro.png}
      \caption{A typical Incident Remediation Pipeline in ITOps.}
      \label{fig:Intro}
    \end{figure}

 \fi 
\section{Related Work}
\textbf{Benchmarking:} 
Traditional coding benchmarks like NL2Bash~\cite{nl2bash2018}, HumanEval~\cite{chen2021evaluatinglargelanguagemodels}, and MBPP~\cite{mbpp2021} treat coding as a sequence transduction task, converting instructions directly into code without interactive execution. Recent efforts have expanded into interactive coding for Python, Bash, and other languages. Previous works~\cite{huang2022executionbasedevaluationdatascience, lai2022ds1000naturalreliablebenchmark,intercode2023} use  Jupyter Notebooks and docker containers as an execution environment to support automatic execution-based evaluation. 

\textbf{Generation and Refinement:}
Recent work on code generation and refinement can be classified into three main approaches: (1) In-Context-Learning ~\cite{learningalgorithmincontextlearning2023, rethinkingroledemonstrationsmakes2022, explanationincontextlearningimplicit2022} enable models to  adapt to new context data at deployment time without requiring traditional fine-tuning or parameter updates;
(2) Chain-of-Thought (CoT) prompting~\cite{chainofthought2023, kojima2023largelanguagemodelszeroshot} enable models to perform multi-step reasoning using internal representations to perform tasks;
%However, CoT reasoning is ungrounded in the external environment, limiting its ability to update knowledge and causing fact hallucination and error propagation.
% To address the limitations of CoT prompting, previous work has been done using ReAct framework and Plan-and-Solve prompting.
(3) ReAct~\cite{react2023} prompts LLMs to generate reasoning traces and actions in an interleaved manner, enabling dynamic reasoning and plan adjustments (reason to act), while also interacting with external environments to refine reasoning (act to reason).

Some of the works like \cite{chen2024teaching, madaan2023selfrefine} have a feedback based framework for the task of code generation and refinement. However, they use unit test cases and execution accuracies for evaluation which makes it hard for adoption concerning Bash use cases.
% Plan-and-Solve~\cite{planandsolve2023} devises a plan to divide the entire task into smaller subtasks, and then carry out the subtasks according to the plan with detailed instructions mitigating calculation, missing step, and semantic misunderstanding errors. 

\textbf{Assessment and Evaluation:}
Recent work on code evaluation can be classified into four main approaches: (1) Match-based: Metrics like BLEU and CrystalBLEU~\cite{papineni2022bleu, eghbali2023crystalbleu} rely on n-gram matching to assess code similarity. (2) Embedding-based: Methods such as CodeBertScore~\cite{zhou2023codebertscore} measure similarity between generated and reference code by using token embeddings and contextual information. (3) Execution-based: These approaches evaluate code quality based on runtime correctness, using metrics like pass@$k$~\cite{kulal2019spoc, chen2021evaluatinglargelanguagemodels}.
(4) Prompt-based: Methods utilize LLMs for pairwise comparison (selecting the better of two codes), single answer grading (scoring a single code), and reference-guided grading (using a reference code if available)~\cite{zheng2023judging, liu2023geval, zhuo2024icescore}. CodeSift~\cite{codesift2024} uses a text-to-text approach, comparing the code’s textual representation with the task description to evaluate correctness.

To summarize, the state-of-the-art methods discussed above have primarily been tested on datasets for languages like Python, Java, C, C++, and C\#. However, these approaches cannot be directly applied to Bash scripts for the ITOps domain, as they rely heavily on execution-based accuracy or unit tests, which are challenging to obtain for Bash data with reliable accuracy. To address this gap, we propose the first end-to-end framework that automates both the generation and assessment of Bash scripts.

\iffalse
\section{Bash Script generation}
In this paper, we propose a novel framework to generate system related scripts that can help in remediate an incident automatically. We explain and focus into the Bash script generation as the scope of the paper. The framework focuses on (i) generation of the bash script using the optimal decoding parameters, as per out experimentation, (ii) evaluation of the generated script automatically, without the need of any execution test-bed in client environment, and (iii) refine the script, if needed, using the explanation or reason generated by observing the incorrectly generated script in the first place.

% \subsection{Script assessment}
% \begin{itemize}
%     \item codeSift
%     \item single grading
%     \item semantic embedding
% \end{itemize}
% \subsection{Script Refinement}
%     \begin{itemize}
%         \item self-refine
%         \item feedback based generation
%     \end{itemize}
\fi

\section{ScriptSmith}

We describe the details of ScriptSmith for the automated generation, assessment and refinement of bash scripts. Our framework (Figure~\ref{fig:framework}) aims to get the correct bash script for each of the recommended action steps. First, it tries to find a matching bash script from the catalogue of past actions. If none is found, it generates a new script dynamically. As the user validates the generated scripts for various kinds of recommended actions, they are added to the catalogue, for future reference. %Figure~\ref{fig:framework} shows our proposed method of \textit{ScriptSmith}.

\begin{table*}[ht]
% \SetTblrInner{rowsep=0pt}
\caption{Accuracy of script generation, assessment, refinement based on Execution. }
\label{tab:result}
\begin{tblr}[]{
    hline{1,2,3,4,Z} = {1pt},
    hline{5-Y} = {solid},
    colspec={X[-1,c,m]|X[-1,c,m]X[-1,c,m]|X[-1,c,m]X[-1,c,m]|X[-1,c,m]X[-1,c,m]},
    rows={font=\scriptsize},
    row{1,2,3} = {font=\scriptsize\bfseries},
    column{1} = {font=\scriptsize\bfseries},
    cell{4}{2,3}={r=2}{c},
    cell{4}{1}={r=5}{c},
    cell{6}{2,3,4,5}={r=2}{c},
    cell{9}{1,2,3,6}={r=2}{c},
    cell{4}{6}={r=2}{c},
    cell{1}{1}={r=3}{c},
    cell{1}{2,4}={c=2}{m},
    cell{1}{6}={c=3}{m},
    cell{2}{2,3,4,5,6,7,8}={r=2}{c},
    cell{2}{7}={c=2}{m},
}

Dataset & Script Generation &  & Script Assessment with CodeSift  &  & Script Refinement &  &  \\
& Model & Accuracy &  Model   &  Accuracy  & Model & Accuracy &  \\
&  & &   &   &  &   \\
Bash Dataset from CodeSift~\cite{codesift2024} & Llama3\_70B           & 75\%              & Llama3\_70B      & 69\%               & Llama3\_70B                           & 75\%(+0)                          &              \\
        &          & 75\%              & Llama3\_8B       &  74\%                  &                                 &      78\%\textbf{(+3)}                         &              \\
           % &          & 74\%              & Gemini1.5\_Flash       & 74\%                   &                                 &  76\%                             &              \\
& Llama3\_8B                      & 46\%              & Llama3\_8B       & 75\%                   &         Llama3\_8B                        &   50\%\textbf{(+4)}                            &              \\
&                      & 46\%              &       & 75\%                   &         Llama3\_70B                        &   63\%\textbf{(+17)}                           &              \\ 
& Gemini1.5\_Pro\footnote{Experiments with Gemini1.5\_Pro model were run on a subset of 50 data points due to the limitation in free API usage.}        & 78\%              & Gemini1.5\_Flash & 84\%               & Gemini1.5\_Pro                           & 84\%\textbf{(+6)}                             \\ 
Bash Dataset from InterCode~\cite{intercode2023} & Llama3\_70B           & 42\%              & Llama3\_70B      & 54\%               & Llama3\_70B                           & 49\% \textbf{(+7)}                         &              \\ 
&            & 42\%              & Llama3\_8B      & 61\%               &                            & 52\%  \textbf{(+10)}                        &              
\end{tblr}
\end{table*}

\subsection{Script Generation using LLMs}
Scripts are generated using LLMs if a similar action statement is not been found in the catalogue. The steps of script generation are as follows:

\begin{enumerate}
    \item 

\textbf{Initial Script Generation -} We generate the script using a code-based LLM. A post-processing step is performed as the raw output of LLMs may have scripts enclosed within other texts. We extract scripts following predefined rules, such as capturing text enclosed in three backticks.

\item
\textbf{Script Evaluation without Execution Bed -} We use the evaluation framework proposed in CodeSift~\cite{codesift2024} to ensure that the generated script aligns with the desired behavior specified by a given task. It involves three main steps: \textit{similarity analysis}, \textit{difference analysis}, and \textit{ensemble synthesis}. The process starts by using syntax checkers to identify any syntactically incorrect script. Next, the framework generates the script functionality and begins the \textit{similarity} and \textit{difference} analysis between the generated functionality and the given task, by prompting on pre-trained LLMs. The final \textit{ensemble synthesis} integrates the \textit{similarity and difference analysis} results to determine the script's functional correctness comprehensively. If either analysis indicates a deviation from the task, the script is labeled as functionally incorrect. 

% It has been observed that the answer from LLM differs if we ask the similarity and dissimilarity between script functionality and task separately. A consensus is obtained when both of the output are being considered, which leads to less number of false positives. It is also to be noted, that CodeSift provides a way to evaluate generated script without execution. 
CodeSift is particularly helpful where it is difficult to write the unit test cases, certain prerequisites are required (eg. move \textit{file1} from \textit{dir1} to \textit{dir2} - \textit{file1}, \textit{dir1} and \textit{dir2} should be present) or there are no absolute answer of a script to match to (eg. free memory in the system).
\item
\textbf{Script Refinement -} If the evaluation step identifies the generated script to be incorrect, we refine the script based on model generated feedback. We first prompt LLMs to briefly explain why the script fails to perform the specified action. We then use this explanation as feedback to prompt LLMs to refine the generated script. %We use a zero-shot prompting mechanism for the script refinement. We tried few-shot prompting techniques but there was no significant improvement in the results.

%Next, we highlight the performance of ScriptSmith, for various settings and analyze the results in details.

\end{enumerate}

Hence, ScriptSmith automatically generates Bash scripts for a given action without human intervention or reliance on an execution environment, thereby enhancing the SRE experience by significantly improving the overall accuracy of script generation.

\section{Results}

In this section, we study the efficacy of ScriptSmith for the automated generation and refinement of Bash scripts using LLMs. Our experimentation primarily centres on the script generation and refinement processes. For script retrieval, we employ state-of-the-art methods, while acknowledging that the current catalog is limited and will expand over time as the deployed system continues to be utilized. The results are summarized in Table~\ref{tab:result}.

% The experiments were designed to evaluate the effectiveness of different models and strategies in improving code accuracy through various methods, including self-debugging, self-refinement, and feedback mechanisms. 

\subsection{Performance of ScriptSmith}
We evaluate the performance of  ScriptSmith on two Bash datasets from CodeSift~\cite{codesift2024} and InterCode~\cite{intercode2023}. For the Bash dataset from CodeSift, which has $100$ samples, we utilize Execution Accuracy (EA) using the testbed provided by CodeSift to determine the correctness of generated and refined script. For the Bash dataset from Intercode consisting of $153$ samples, we ask the domain experts to evaluate the correctness of the generated and refined script. This is due to the unreliability of the execution environment provided by InterCode as discussed in the User Study ~\ref{user_study} section.

We compare the performance of script generation, assessment, and refinement across four models: Llama3\_8B, Llama3\_70B, Gemini1.5\_Flash, and Gemini1.5\_Pro. 
We primarily explore two different configurations of script generation/refinement and script assessment models: 1) \textit{Self-Reflection}: Both script generation and script assessment models are the same. 2) \textit{Peer-Review}: A smaller model is used to evaluate the script quality generated by a larger model. The motivation for peer review is that models are often biased when evaluating their own generated output.
For example, when evaluating scripts generated by Llama3\_70B, CodeSift's assessment accuracy using Llama3\_8B increases to $74\%$ from $69\%$ for CodeSift-Bash Dataset and $61\%$ from $54\%$ for Intercode-Bash Dataset when compared to Llama3\_70B. This indicates that using peer review yields better results than using self-reflection. Furthermore, we select a larger model for script generation and a smaller model for script assessment to reduce costs, as script assessment requires more LLM calls (and tokens) than script generation.  

The performance of script assessment also affects script refinement performance. For Llama3\_70B model, we see that assessment with Llama3\_8B model (peer-review) results in $3\%$ and $10\%$ improvement in script accuracy for CodeSift and Intercode dataset respectively compared to $0\%$ and $7\%$ when assessment is done with Llama3\_70B model (self-refine).

We also compare the performance of open-source and closed-source models. As can be seen from Table \ref{tab:result}, the closed-source Gemini1.5 model outperforms the open-source Llama3 model by $6\%$ on the CodeSift-Bash dataset. However, cost of calling gemini1.5 models is much higher than Llama3 models (which can be run locally). 

Finally, we explore another configuration where we keep script generation and script assessment models as Llama\_8B (smaller sized model) but change the script refinement model to Llama\_70B. The motivation behind this configuration is that the number of calls to the LLM is much less in the script refinement phase as compared to the script generation and assessment phase as it is only applied to instances flagged as incorrect during assessment. In this configuration, we observe the greatest improvement in script refinement accuracy—17\% in the CodeSift-Bash dataset.

To summarize, we have the following takeaways from our experiments:
\begin{itemize}
    \item Accuracy of generated scripts increases using ScriptSmith framework for bash scripts in ITOPs domain. The increase is significantly bigger when initial script generation accuracy is less. However, if the initial accuracy is high, then refinement does not add significant value due to the saturation of model performance.
    \item Peer-Review performs significantly better than Self-Refine since it does not suffer from biases.
    \item Performance of open-source models with ScriptSmith (through automatic assessment and refinement) can match the performance of raw closed-source models for script generation.
\end{itemize}

\section{Deployment}
 \begin{figure}[!htp]
      \centering
      \includegraphics[width=1.0\linewidth]{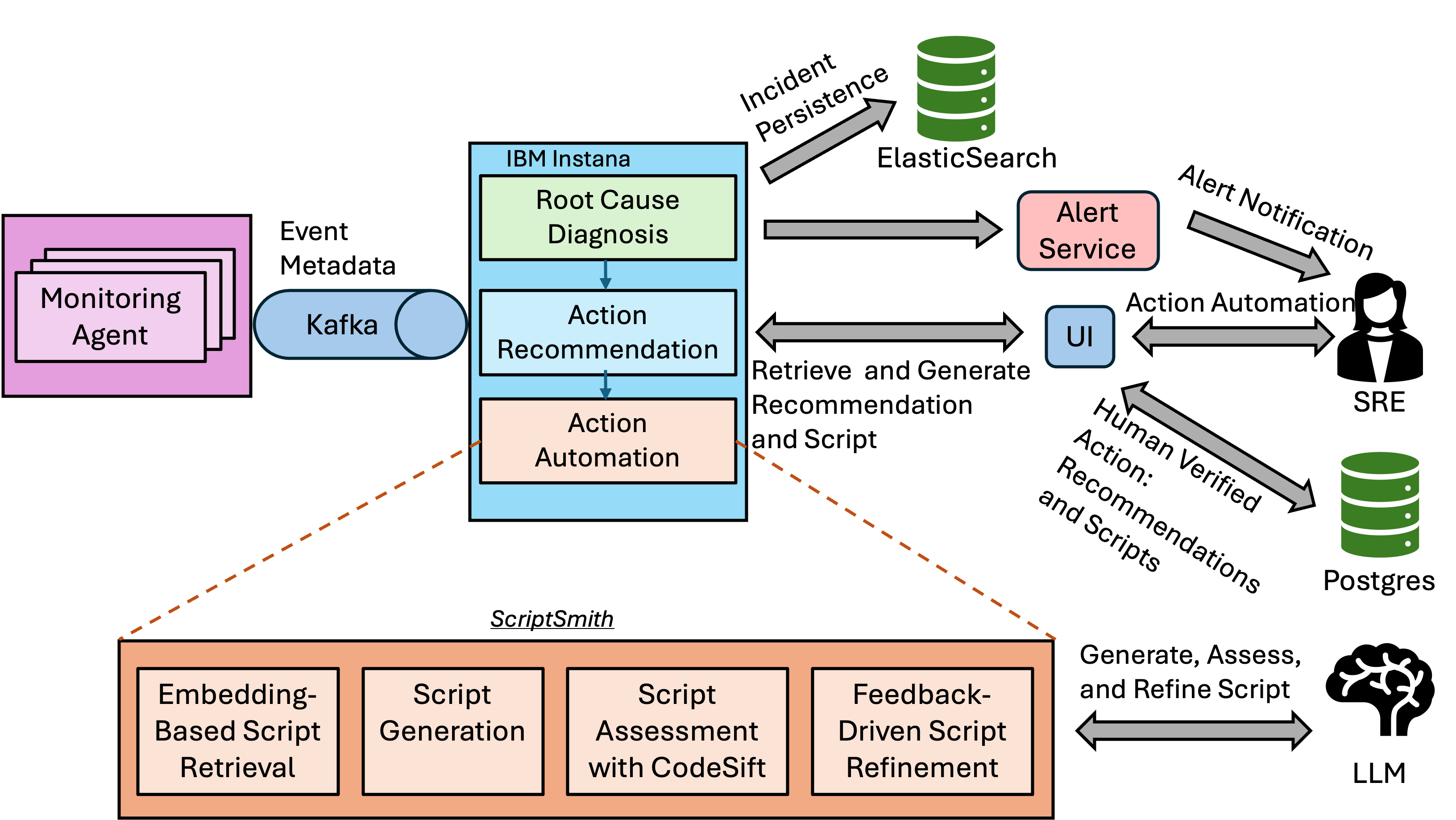}
      \caption{IBM Instana's Intelligent Remediation Deployment Pipeline with ScriptSmith}
      \label{fig:arch}
    \end{figure}
Figure~\ref{fig:arch} shows the intelligent remediation pipeline's complete software architecture, including the automation generation block (ScriptSmith) and its modules. 
The monitoring agents deployed in the user environment collect observability data through policies\footnote{https://www.ibm.com/docs/en/instana-observability}. 
These event metadata are pushed to IBM Instana from Apache Kafka. 
The Incident Processing Service combines the event metadata from different monitoring agents and creates an incident report for them.  
After the incident is stored in the ElasticSearch database, an alert is created by the Alert Service to notify SREs via Slack or Pagerduty\footnote{https://www.ibm.com/docs/en/instana-observability/current?topic=instana-managing-events-alerts}.
An SRE logs into the Instana UI to diagnose the issue with the help of the root cause diagnosis service and works to mitigate and remediate the incident.
SRE first uses the Action Recommendation Service to create human-assisted steps. SRE then triggers the action automation framework to recommend bash scripts for a specified action that is divided into two primary steps:
\begin{itemize}

\item The framework first attempts to retrieve a relevant script from its pre-existing knowledge catalog. This catalog is a repository of verified scripts that the model has previously encountered and the solution of which has been stored.
We build an embedding database by converting all script descriptions in our catalogue into high-dimensional vectors using transformer models. These vectors are indexed for efficient similarity searches using approximate nearest-neighbor algorithms. When our tool receives a prescription text, it transforms it into a vector, retrieves the most similar vectors from the indexed database, maps these vectors back to their original script descriptions, and then returns the relevant scripts. If the model can retrieve a script with high confidence, this script is directly shown to the user. The confidence level is determined by the model’s similarity measures and relevance scoring against the user’s request.
\item For cases where the framework cannot retrieve a script from the knowledge catalogue, it generates a new script using LLMs. The framework incorporates an assessment step before presenting the code to the user. We use the approach presented in ~\cite{codesift2024} for evaluating the generated script without any execution environment. If the validation identifies the script as incorrect, the model is prompted to explain why the script is wrong given the incident. This explanation is then used to regenerate the script, to provide the user with the correct script for incident remediation.
%She can refine the recommended steps before triggering the Action Automation Service.
\end{itemize}

Finally, each recommended script (either retrieved or regenerated) is reviewed by a SRE for its correctness.  Based on their domain knowledge, the SRE reviews the script, approves it, makes changes, or rejects the recommendation entirely. The final recommendation is then published in a Postgres database serving as our curated knowledge catalog, enriching the catalog with verified and improved scripts. This continuous feedback loop ensures that the knowledge catalogue evolves and improves over time, reducing the need for frequent script generation and enhancing the accuracy and relevance of  script recommendations. The framework is designed to prioritize script quality and minimize noise in recommendations. By leveraging the dual approach of retrieval and generation, along with built-in validation and feedback mechanisms, the system ensures that users are presented with scripts that are both functional and relevant to SRE's needs. This method streamlines script generation and refines script quality through continuous learning and validation.

The framework for internal user study has been running on Instana for the last six months. The user study, described in the next section, turned out to be highly useful, especially in the present conditions where the lack of adequate ground truth and execution test environment inhibits proper performance evaluation. This framework has led to the creation of the knowledge catalogue, and feedback collection and in turn helping the LLMs to improve. The proposed framework is being deployed as a tech-preview to assist SREs in effective and faster remediation of various incidents. The integration of the proposed framework with Instana enables SREs to evaluate the recommended scripts and provide feedback in real-time. 

% Figure~\ref{fig:ui} shows a snapshot of the Instana's user interface where the user can key in the action prescription. The recommended scripts, code/script usage, and semantic validation are reflected on the UI.

\iffalse
     \begin{figure}[!htp]
      \centering
      \includegraphics[width=1.0\linewidth]{CameraReady/LaTeX/figs/UI.png}
      \caption{User Interface for Bash action automation}
      \label{fig:ui}
    \end{figure}

\fi
\section{User Study} \label{user_study}

\begin{table*}[ht]
\caption{Examples Illustrating Key Aspects of Code Generation Evaluation: Human Judgment, Execution-Based Assessment, and Model Feedback Effectiveness.}
\label{tab:user_study}
\scriptsize
\centering
\begin{tblr}{
        hline{1,2,Z}={1pt},
        hline{6,10,14,18}={solid},
        colspec={X[-1,l]|X[2,l]|X[-1,l]|X[-1,l]|X[-1,l]},
        rows={m,rowsep=0pt},
        row{1}= {c},
        cell{2,6,10,14,18}{1,2,3,5}={r=4}{l},
    }
\textbf{Input Task}       & \textbf{Generated Script}   & \textbf{Outputs} & \textbf{Assessment} & \textbf{Model Feedback} \\

Print disk usage of files or folders in /workspace
&
\begin{minipage}{0.7\columnwidth}
\textbf{First Pass Code:}
\begin{minted}{bash}
#!/bin/bash
for file in /workspace/*; do
    du -sh "\$file"
done
\end{minted}
\end{minipage}
&
\begin{minipage}{0.4\columnwidth}
\textbf{Expected:}
\begin{minted}{bash}
N/A
\end{minted}
\textbf{Actual:}
\begin{minted}{bash}
N/A
\end{minted}
\end{minipage}
& \textbf{EA:} \xmark
& N/A \\
& & &\textbf{CodeSift ($70b$):} \cmark &\\ 
& & &\textbf{CodeSift ($8b$):} \cmark &\\ 
& & &\textbf{Human:} \cmark &\\ 

Save first IP address of domain `google.com' in `address' variable and display it.
&
\begin{minipage}{0.7\columnwidth}
\textbf{First Pass Code:}
\begin{minted}{bash}
#!/bin/bash
address=$(dig +short google.com | head -1)$
echo "The first IP address of google.com is: $address"
\end{minted}
\end{minipage}
& 
\begin{minipage}{0.4\columnwidth}
\textbf{Expected:}
\begin{minted}{bash}
142.250.195.206
\end{minted}
\textbf{Actual:}
\begin{minted}{bash}
The first IP address of google.com is 142.250.195.206
\end{minted}
\end{minipage}
& \textbf{EA:} \xmark
& N/A \\
& & &\textbf{CodeSift ($70b$):} \cmark &\\
& & &\textbf{CodeSift ($8b$):} \cmark &\\ 
& & &\textbf{Human:} \cmark &\\ 

Check if current shell is running within a 'screen' process.
&
\begin{minipage}{0.7\columnwidth}
\textbf{First Pass Code:} 
\begin{minted}{bash}
#!/bin/bash
if [ -n "\$STY" ]; then
    echo "Running within a screen session"
else
    echo "Not running within a screen session"
fi
\end{minted}
\end{minipage}
& 
\begin{minipage}{0.4\columnwidth}
\textbf{Expected:}
\begin{minted}{bash}
1
\end{minted}
\textbf{Actual:}
\begin{minted}{bash}
Running within a screen session
\end{minted}
\end{minipage}
& \textbf{EA:} \xmark
& N/A \\
& & &\textbf{CodeSift ($70b$):} \cmark &\\
& & &\textbf{CodeSift ($8b$):} \cmark &\\ 
& & &\textbf{Human:} \cmark &\\ 

Copies all files with `FooBar' in the path under /testbed/dir1 directory to /testbed/dir3/subdir1 /subsubdir1/tmp directory. &
\begin{minipage}{0.7\columnwidth}
\textbf{First Pass Code:}
\begin{minted}{bash}
#!/bin/bash
find /testbed/dir1 -type f -name "*FooBar*" -exec cp {} /testbed/dir3/subdir1/subsubdir1/tmp \;
\end{minted}
\textbf{Refined Code}:
\begin{minted}{bash}
#!/bin/bash
find /testbed/dir1 -type f -path "*FooBar*" -exec cp {} /testbed/dir3/subdir1/subsubdir1/tmp \;
\end{minted}
\end{minipage}
&
\begin{minipage}{0.4\columnwidth}
\begin{minted}{bash}
N/A
\end{minted}
\end{minipage}
&
\textbf{EA:} \xmark
&
The generated code is incorrect because it will copy all files whose name contains `FooBar', not all files with `FooBar' in the path. To fix this, the `-name' option should be replaced with `-path'. \\
& & &\textbf{CodeSift ($70b$):} \xmark &\\ 
& & &\textbf{CodeSift ($8b$):} \xmark &\\ 
& & &\textbf{Human:} \xmark &\\ 

Find files in the /workspace directory and sub-directories, that were accessed within last hour. &
\begin{minipage}{0.7\columnwidth}
\textbf{First Pass code:}
\begin{minted}{bash}
#!/bin/bash
find /workspace -type f -amin -60 -print
\end{minted}
\textbf{Refined Code:}
\begin{minted}{bash}
#!/bin/bash
find /workspace -type f -mmin -60 -print
\end{minted}
\end{minipage}
& 
\begin{minipage}{0.4\columnwidth}
\begin{minted}{bash}
N/A
\end{minted}
\end{minipage}
&
\textbf{EA:} \cmark
& It uses `-amin' option which stands for `access time in minutes'. Since the task requires to find files accessed within the last hour, we should use `-mmin' (modified time in minutes). \\
& & &\textbf{CodeSift ($70b$):} \xmark &\\ 
& & &\textbf{CodeSift ($8b$):} \xmark &\\ 
& & &\textbf{Human:} \cmark &\\ 
\end{tblr}
\end{table*}

In our study, we involved four domain experts to evaluate the performance of the Bash script generation model. We asked the experts to label the initial script, the model-generated feedback (explanation for the error), and the refined script. Experts labeled instances on a scale of 0 (incorrect), 1 (partially correct), and 2 (correct). We used two criteria: strict (only 2 is correct) and partial (1 or 2 is correct). The goal was to compare the accuracy of the model’s initial and refined output and assess the usefulness of the feedback provided to fixing the bugs. 
The experts provided feedback for $153$ cases from the interCode Bash dataset ~\cite{intercode2023}. The script generation accuracy for the first pass was $42\%$ using \textit{strict} labeling criteria. The cases that were identified as incorrect (labeled as either $0$ or $1$) were refined using the model's feedback. This resulted in an overall accuracy of $76\%$. %under strict labeling criteria, showing that the refinement model improved overall accuracy by 22 percentage points. When considering partial correctness, the first pass accuracy was 53\%, which increased to 73\% after refinement, reflecting a 20 percentage point improvement.

% The evaluation focused on three key areas: (1) the alignment of automatic code assessment with human judgment,  (2) the effectiveness of the automatic feedback provided by the model for refining the code, and (3) the accuracy of the final refined code.
From the user study, we analyze the following four key aspects: %(1) Alignment is human judgement and execution-based evaluation, (2) the alignment of automatic code assessment with human labelling, (3) the effectiveness of the model’s automatic feedback in refining the code, and (4) the usefulness of the proposed framework. Below, we discuss each of these findings in detail.

\begin{itemize}
\item \textbf{Expert Judgment vs. Execution Accuracy}:
% We examine the alignment between human judgment and execution accuracy (EA). 
Expert judgment shows an initial script generation accuracy of $42\%$, while EA reports $27\%$. We perform a detailed analysis to understand this discrepancy and identify three primary reasons:
\begin{enumerate}
    \item \textit{Different Interpretations:} Expert evaluators and the execution-based system can interpret the input task differently, leading to varying assessments of script correctness.  Row 1 in Table~\ref{tab:user_study} illustrates how divergent interpretations resulted in different evaluations. The execution environment expects disk usage of files and folders in subdirectories where as human is satisfied with disk usage of file only in the given directory.

\item \textit{Restricted Execution-Based Evaluation}: EA's critiques are too stringent to be considered fair. In row 2 in Table~\ref{tab:user_study}, the script's additional text output alongside the IP address led to a misleading assessment, as the execution environment required only the IP address. Similar issues arise when the EA expects precise final answers, and accompanying text causes the script to be incorrectly labeled.

\item \textit{Incorrect Expected Output}: There were also cases where the expected output for the given task was incorrect. Row 3 in Table~\ref{tab:user_study} has is an example of such a case where the exepected output is 1 (checking the number of processes) instead of boolean answer whether current shell is within a screen process.

\end{enumerate}
Given these discrepancies, we decided to rely on expert judgment to analyze the other aspects of the study. This approach ensured a more accurate and consistent evaluation of the model's performance and the effectiveness of the ScriptSmith framework.

   \item \textbf{Expert Judgement vs. CodeSift Assessment:} Next, we examine the alignment between automatic script assessment (CodeSift) and expert preferences, using two models for evaluation.
   CodeSift's assessment using the $Llama3\_8b$ model matched expert annotations in $61\%$ of the $153$ cases under \textit{strict} labeling criteria and in $66\%$ of the cases under \textit{partial} labeling criteria. In comparison, with the $Llama3\_70b$ model, CodeSift showed a lower alignment with expert annotations, with $54\%$ for \textit{strict} labeling and $63\%$ for \textit{partial} labeling. These results suggest that the larger model, $Llama3\_70b$, may exhibit self-bias, particularly in cases where it incorrectly labels script as correct as shown in row 3 in Table~\ref{tab:user_study}.
   
%  \item  \textbf{Overall Accuracy and Alignment:} \\
%  Here the objective is to check the alignment of the execution based accuracy with human preference. 
% \textbf{First Pass}: Out of xx cases, 47 were labeled as incorrect (0) by the execution-based evaluation. When experts reviewed these cases, they found that the model's initial assessment sometimes aligned better with human judgment than with the execution bed results. Specifically, the experts found that out of the 47 cases marked as incorrect by the execution tests, 12 were actually correct upon manual review. This indicates that the execution environment is rigid with specific configurations and setups, which may lead to some valid scenarios being mistakenly identified as incorrect. \red{should we fix the 12 cases?}

% \textbf{Refined Code}: After applying the model’s automatic feedback, 35 out of the 47 initially incorrect cases were corrected. This improvement shows the model's potential in refining code through feedback, though a small percentage of cases still required further human intervention. 

 \item  \textbf{Usefulness of Model-Generated Feedback:}
We assess the effectiveness of model-generated feedback in two ways: (1) \textit{Human Support}, i.e., computing the frequency of cases where experts found the feedback to be useful, and (2) \textit{Model Correction}, i.e., computing the frequency of cases where the model used the feedback to correct the script. For this analysis, we applied \textit{strict} labeling criteria. Among the $88$ cases that experts labeled as incorrect during the first pass, they reviewed the reasons provided by the model for the script's incorrectness. In $69\%$ of these cases ($61$ out of $88$), experts found the feedback to be correct.  Additionally, in $77\%$ of the cases where the feedback was labeled as $2$ (correct), the model was able to use this feedback to successfully correct the script. Row 4 in Table~\ref{tab:user_study} illustrates a scenario where the feedback generated by the model accurately identifies where the generated script goes wrong and suggests a specific command to fix the error, resulting in accurate script refinement. %We discovered that the model successfully refined incorrect code 56\% of the time. Among these cases, experts found the model-generated feedback to be useful 88\% of the time. In cases where the model failed to refine the code, we observed that 65\% of the time, experts labelled the feedback as either incorrect or partially correct. This suggests that, in the majority of scenarios, the LLMs are capable of reasoning and refining the code effectively. \red{This statement does not follow from the numbers}
%Interestingly, there were scenarios where the model's feedback was labelled as 0 (incorrect), yet the refined code turned out to be correct. This indicates that even when the feedback does not appear directly useful, the model may still leverage the signal that the initial code was incorrect to improve the code. \red{We should probably skip this statement. Else we need to say why this happens.}
There were very few instances (less than $4\%$) where even though the feedback was labeled as $1$ (partially correct) or $2$ (correct) but the model failed to refine the script successfully. \textbf{\textit{Observation: The model struggles to consistently adapt feedback for refinement if they are verbose.}} Overall, the feedback helped SREs save time during the debugging process.
 \textbf{\textit{Recommendation: Additional steps may need to be introduced in the pipeline to incorporate feedback.}}. In some cases, when experts provided specific reasons for the initial script being marked as incorrect, the model was able to refine the script effectively after receiving targeted feedback. 

% \textbf{Expert Evaluation}: In 77\% of the cases, experts found the model-generated feedback to be useful for SREs in identifying and correcting bugs. However, there were instances where the feedback was labeled as 1 (partially correct) or 2 (correct), but the model failed to refine the code successfully. Despite this, the feedback itself proved valuable, helping SREs save time during debugging. \\
% \textbf{Cases of Misalignment}: 
 \item  \textbf{Effectiveness of the Proposed Framework}:
We assess the usefulness of the proposed framework, specifically by considering the automatic script assessment using CodeSift with the $LLama3\_8b$ model. The scripts labeled as incorrect by the CodeSift model were then considered for refinement. Out of $153$ cases, CodeSift correctly identified $38$ incorrect cases, and out of these, $47\%$ were successfully corrected using automatic feedback (considering \textit{strict} correctness criteria) and $62\%$ in case of partial correctness. However, we also encountered scenarios where initially correct scripts were wrongly assessed as incorrect by CodeSift. \textbf{\textit{Observation: During refinement, asking the model to explain errors led to hallucinations, turning a correct script into an incorrect one.}}
%During the refinement process, despite the script being correct, the \textbf{\textit{instruction to explain why the script was wrong led the model to hallucinate reasons for the error}}, resulting in an incorrect script after refinement. 
Row 5 in Table~\ref{tab:user_study} illustrates such a scenario, where the model's unnecessary attempts to identify errors in a correct script led to incorrect final output. \textbf{\textit{Recommendation: Add guardrails during prompting to prevent the model from self-doubt.}} Overall, we observed a $10\%$ improvement in the accuracy of the script generation pipeline. The proposed framework uses automation to improve script accuracy, enhance the expert experience, and streamline workflow by reducing manual debugging and refinement time.

%Our analysis of the proposed framework reveals a significant improvement in code generation accuracy. \red{Proposed framework is with automatic evalution. So we should not write 10\% here and not 22.} Based on human evaluation, there is an overall 22 percentage point increase in the accuracy of the final code compared to the first-pass code. This improvement highlights the effectiveness of the framework in refining code through iterative feedback and refinement.

%However, when relying solely on the automatic code assessment process, we observe that the model identifies only xx out of yy cases as incorrect. Of these xx cases, the model successfully refines zz cases. This results in a 10\% improvement in overall code generation accuracy.
%The alignment between the execution-based evaluation and expert labels improved after refinement, though some discrepancies remained due to the limitations of the execution bed. \\

%In certain cases, the model incorrectly identified that the first-pass code was inaccurate. However, when prompted to regenerate the code, the model occasionally produced incorrect code. This suggests a potential issue with the model being misguided in scenarios where the initial code is already correct.
%\red{Probably need to revisit this section after the numbers.}

\end{itemize}
% The study highlights both the strengths and limitations of using automatic assessment and feedback for code refinement. While the model often improves the accuracy of the code, there are still scenarios where human judgment is essential, particularly in cases where the execution-based evaluation is not fully aligned with expert assessment. The feedback generated by the model is generally useful for SREs, contributing to more efficient debugging and code refinement processes.
\section{Conclusion and Future Work}

In this paper, we introduce ScriptSmith, a reference and execution-free framework for generating Bash scripts. The framework effectively identifies faulty scripts and provides detailed reasoning for these inaccuracies, which in turn helps refine the scripts. %and assist domain experts when the model cannot fully refine them.
Our findings demonstrate that automatically generated feedback improves debugging and helps experts quickly locate and fix issues in the script. The alignment between generated feedback and expert judgment further underscores the potential of this approach to improving script quality in automated settings. %While the current framework shows promise, several areas for improvement and expansion have been identified. 
A key challenge is scaling the testing of generated scripts. This requires developing methods to automatically generate comprehensive test cases that cover a wide range of scenarios, ensuring more robust script validation. Additionally, executing these scripts within a controlled environment would offer more reliable assessments, minimizing discrepancies between execution-based evaluations and expert judgment. In the future, we aim to enhance the effectiveness and reliability of the proposed framework, making it a more valuable tool for automated script generation and refinement. We also plan to explore ScriptSmith to other scripting languages like Powershell.

\bibliography{aaai25}

\end{document}